\documentclass[twocolumn,         % Format : preprint, twocolumn
               showpacs,longbibliography,            % Pacs : showpacs, noshowpacs
               showkeys,preprintnumbers,     % Preprint: preprintnumbers,
               aps,                 % Society: ...
               prd,          	    % Journal Style : pra, prb, prc, prd, pre,
               letterpaper,             % Size : a4paper, ...
               nofootinbib,         % Footnote: footinbib, nofootinbib
               tightenlines,        % Remove additional spaces in a line
               floats,floatfix      % Floating pictures and tables
               ]{revtex4-1}
\usepackage{physics}
\usepackage{epsf}
\usepackage{graphicx,color}
\usepackage{subfigure}
\usepackage{latexsym}
\usepackage{amsmath,amssymb}        
\usepackage[colorlinks=true,linkcolor=blue,citecolor=blue]{hyperref}
\usepackage{mathrsfs}
\usepackage{comment}
\usepackage{soul}
\usepackage{cancel}
\definecolor{purple}{rgb}{0.58,0.0,0.83}
\definecolor{orange}{rgb}{1,0.5,0}
\DeclareSymbolFontAlphabet{\mathrsfs}{rsfs}
\DeclareMathAlphabet{\mathcal}{OMS}{cmsy}{m}{n}

\begin{document}

% -----> TITLE 

%\title{Effects of domain topology on Fuzzy Dark Matter dynamics}
%\title{Circular Orbit Breakdown in Time-Evolving Fuzzy Dark Matter Systems / Dynamics of Test Particles in Anisotropic, Time-Dependent Fuzzy Dark Matter Structures / 
\title{Unveiling Orbital Chaos: The Wild Heart of Fuzzy Dark Matter Structures}

% ----->   AUTHORS   <-----

\author{Iv\'an  \'Alvarez-Rios}
\email{ivan.alvarez@umich.mx}
\affiliation{Instituto de F\'{\i}sica y Matem\'{a}ticas, Universidad
              Michoacana de San Nicol\'as de Hidalgo. Edificio C-3, Cd.
              Universitaria, 58040 Morelia, Michoac\'{a}n,
              M\'{e}xico.}

\author{Francisco S. Guzm\'an}
\email{francisco.s.guzman@umich.mx}
%\thanks{Corresponding author}
\affiliation{Instituto de F\'{\i}sica y Matem\'{a}ticas, Universidad
              Michoacana de San Nicol\'as de Hidalgo. Edificio C-3, Cd.
              Universitaria, 58040 Morelia, Michoac\'{a}n,
              M\'{e}xico.}  

% --->   DATE

\date{\today}

% -----> ABSTRACT

\begin{abstract}
In this paper we study the behavior of test particles on top of a galactic-type of Fuzzy Dark Matter (FDM) structure, characterized by the core-halo density profile found in simulations. Our workhorse structure is an anisotropic, time-dependent, virialized core-tail FDM clump resulting from a multicore merger. For our analysis we allow this structure to keep evolving, which implies that the core oscillates and accretes matter from the halo, while the halo dynamics is dominated by its characteristic high kinetic energy. On top of this time-dependent structure that in turn has a time-dependent gravitational potential, we solve the motion equations of test particles with initial conditions associated to circular orbits at different radii. Our results indicate that: 1) no trajectory remains circular, 2) the trajectories are sensitive to initial conditions and 3) the departure of initially near trajectories has always a positive Lyapunov exponent. A qualitative result is that the motion of test particles is more erratic with a bigger Lyapunov exponent within and near the core than in the halo region, which can be understood in terms of the random motion of the core within the core-halo structure. We expect these results warn on the importance of the anisotropic and time-dependent nature of FDM clumps when studying the motion of test particles.
\end{abstract}

% ----->   PACS

\keywords{Dark Matter -- Bose Condensates}
%\pacs{keywords: self-gravitating systems -- dark matter -- Bose condensates}
%07.05.Tp Computer modeling and simulation
%07.05.Mh Neural networks, fuzzy logic, artificial intelligence
%05.45.Tp Time series analysis
%04.30.-w Gravitational waves

% ----->   MAKETITLE   <-----

\maketitle

% ---------------------------------------------
% ----->     INTRODUCTION.    <-----
% ---------------------------------------------
\section{Introduction}
\label{sec:GalacticStructure}

The Fuzzy Dark Matter (FDM) model postulates that Dark Matter (DM) consists of a gas made of bosonic particles with an ultralight mass of order $10^{-23}-10^{-21}$eV. This model is an alternative to the Cold Dark Matter (CDM) with particular properties, including a similar to CDM large scale behavior, while wave-like properties arise at galactic halo scales \cite{Chavanis2015,Niemeyer_2020,Hui:2021tkt,ElisaFerreira}. The model is being actively studied and recently a number of results indicate particular fingerprints that  will allow a systematic contrast with observations \cite{RevModPhys.93.015004}, for example at solar system \cite{Tsai2023,An_2024} and Earth scales \cite{PhysRevLett.133.251001}, 
at laboratory scale \cite{PhysRevLett.131.011001,PhysRevLett.115.011802,PhysRevLett.123.141102,Bloch_2023} 
at compact object scale \cite{PhysRevLett.119.221103}, using gravitational lensing observations \cite{PhysRevLett.125.111102} and  gravitational wave detectors \cite{Vermeulen_2021,PhysRevLett.133.101001,PhysRevResearch.1.033187}, are efforts and models aiming the search for particular signals of this dark matter candidate.

The model is not free of debatable essential properties, for example the boson mass faces a number of constraints arising from varios observation \cite{PhysRevLett.122.231301}, for example in \cite{dalal2022fuzzy,PhysRevLett.123.051103,Hayashi_2021} constraints from Milky Way Satellite are studied, also from Ultrafaint Dwarf Galaxies \cite{2022PhRvD.106f3517D}, from stellar motion near SMBHs \cite{Della_Monica_2023,Della_Monica_2023b} and from Lyman-$\ensuremath{\alpha}$ constraints   \cite{PhysRevLett.119.031302,PhysRevLett.126.071302}. While these are mass constrains from different observations, in our analysis below we use the boson mass $10^{-23}$eV because it has been shown to fit rotation curves of dark matter dominated dwarf galaxies (e.g. \cite{Tula2018}).

The bosons of the FDM are assumed to be in a coherent state, characterized by the macroscopic wave function $\Psi$ known as the order parameter. The potential trap for the FDM is the self-generated gravitational potential $V$, sourced by its own gas density distribution. The dynamics of the order parameter in the FDM model is governed by the Schr\"odinger-Poisson (SP) system:

\begin{eqnarray}
i\hbar \dfrac{\partial \Psi}{\partial t} &=& -\dfrac{\hbar^2}{2 m_B}\nabla^2\Psi + m_B V \Psi, 
\label{eq:GP}\\
\nabla^2 V &=& 4\pi G \left(\rho - \bar{\rho}\right),
\label{eq:Poisson}
\end{eqnarray}

\noindent where $\rho = m_B |\Psi|^2$ is the bosonic gas density, and $\bar{\rho}$ is the spatial average density over the  domain of solution. Interesting FDM structures are characterized by a core-halo density profile discovered in the seminal structure formation simulations \cite{Schive:2014dra}. One first spectacular result of such analysis is that the core has a density profile that in average -both, spatial and time averages- is similar to that of the ground state solution of the SP system constructed assuming isolation boundary conditions \cite{GuzmanUrena2004}, while a second one is that the halo profile could be fitted with the NFW profile \cite{NFW,IvanTulaChavanis2024}.

The core-halo profile of FDM structures seems to be the elementary brick in the cosmic web of structures and therefore its study is essential to the model. Studies at local scale include formation history related scaling relations (e. g. \cite{Schive:2014hza,Schwabe:2016,Mocz:2017wlg,DuNiemeyer2017,IvanFranciscoCoreTailSols,Chan_2022}), as well as attempts to show that these configurations can be useful at fitting for example galactic rotation curves using oversimplified density models (e.g. \cite{Bernal:2017oih}). 

The construction of density profiles is not free of subtleties. In structure formation simulations of CDM from which NFW profiles are obtained \cite{NFW}, as well as simulations of structure formation of FDM (e.g. \cite{Schive:2014dra,Mocz:2017wlg,Veltmaat_2018,May_2021,Gotinga2022}) an overdensity is located and at that point a spherically symmetric density profile is constructed using an average over angular directions. This is the reason why, for example, the NFW density profile of CDM and the universal density core of FDM are written as function of the spherical coordinate $r$, of a reference system centered at the overdensity. The real density profile of an FDM structure can in fact be approximated with a multimode expansion as described in \cite{PhysRevD.97.103523}, and is not spherically symmetric at all, instead, it has a profile that highly depends on the angular directions at every snapshot of a simulation.

Moreover, FDM structures are time-dependent and density profiles are also the result of an additional time-average over the already solid angle averages. Thus, when a formula for the core-halo density profile is given as function $r$ only, it refers to a spatial and temporal average, it does not mean that the density is spherically symmetric and stationary. Instead, it has multipoles of high order and has a time dependence given by the time-dependent system (\ref{eq:GP})-(\ref{eq:Poisson}).  

Some of the dynamical properties of these anisotropic non-stationaryFDM structures at small scales have been analyzed. These include the study of the stochasticity of  fluctuations of FDM \cite{Centers_2021} and the random walk behavior of the core within an FDM halos \cite{Li_2021,PhysRevLett.124.201301,Dutta_Chowdhury_2021}, which may lead to possible interesting observable effects. We then go a step in this direction and investigate a basic implication of the stochastic behavior of FDM at core-halo scales.
The spatial and time dependency of the FDM core-halo structure density, implies that the gravitational potential is also anisotropic and time-dependent. We then explore the motion of test particles beyond the idealized spherically symmetric universal averaged density profiles found for FDM.
The objective of our analysis is to study the motion of test particles subject to the gravitational potential of one of these formed core-halo structures, under the assumption that the core-halo configuration is anisotropic, time-dependent and evolves according to Eqs. (\ref{eq:GP})-(\ref{eq:Poisson}).

The paper is organized as follows. In Section \ref{sec:corehalo} we describe the workhorse core-halo structure of FDM that we use for our analysis. In Section \ref{sec:eqmotion} we describe the motion equations of test particles and in Section \ref{sec:soleqmotion} we present the results. Finally in Section \ref{sec:comments} we draw some conclusions.
% -------------------------
\section{Stationary granular Core-Halo}
\label{sec:corehalo}

\subsection{The workhorse core-halo structure}

As mentioned before, the formation of these structures was first discovered in structure formation simulations 
\cite{Schive:2014dra,Mocz:2017wlg,Veltmaat_2018,MoczPRL2019,May_2021,Gotinga2022}. However,  soon later on they were constructed under less computationally demanding scenarios, specifically via the multimerger of cores as described in \cite{Schive:2014hza} and later works \cite{Mocz:2017wlg,Schwabe:2016}. Going beyond, core-halo structures were recently constructed {\it ab initio}, without the need of evolution of structures or multi-solitons, but through the multimode expansion of $\Psi$ as described in \cite{YavetzLiHui2022} and reanalyzed in \cite{IvanTulaChavanis2024}.

\begin{figure}
\includegraphics[width=8cm]{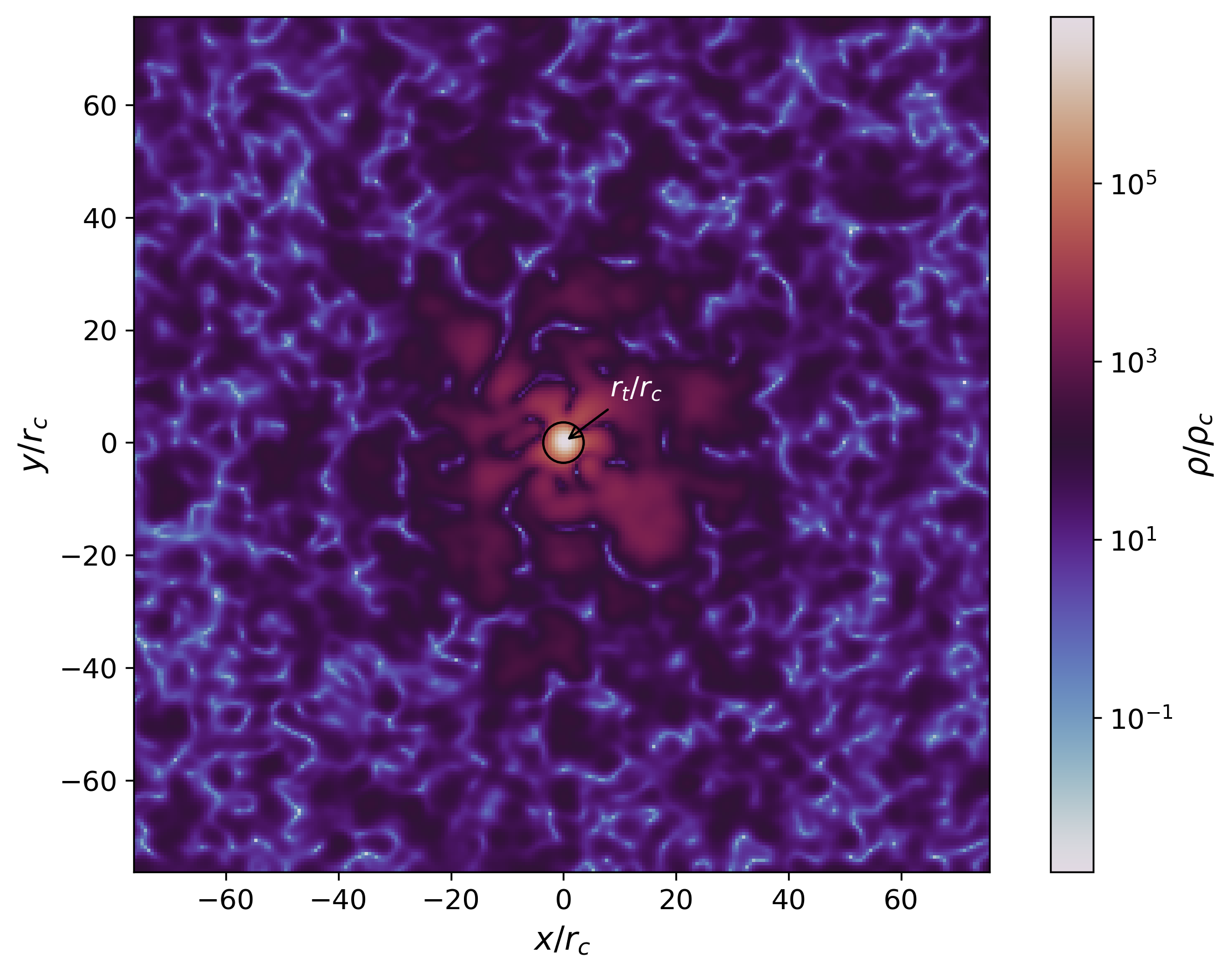}
\caption{Density of the workhorse core-halo configuration projected on the  $z=0$ plane, which results from the evolution of a multicore merger.  The axes are normalized with respect to the core radius $r_c\approx 0.2621$ kpc, while the density is normalized with respect to  $\rho_c \approx 4.201\times 10^{11}\text{ M}_\odot / \text{kpc}^3$. The circle represents the transition radius $r_t/r_c \approx 3.597$ between the core and its envelope. This numerical domain is the laboratory reference frame used in our analysis below.}
\label{fig:initialDensity}
\end{figure}

From these three methods, we use the structure formed from a multicore merger that has undergone relaxation as described in \cite{periodicas}, whose density is shown in Figure \ref{fig:initialDensity}. This is the result of the merger of 10 soliton cores with random masses that resulted in a final mass of $\sim 2.598\times10^{11}$ M${}_\odot$ evolved using the code CAFE-FDM  \cite{Alvarez_Rios_2022,periodicas} using a boson mass $m_{B}=10^{-23}$eV. These cores are initially at rest, randomly located within a cubic domain of size 40 kpc, which in turn is discretized with a resolution of $h = 5/32$ kpc. The evolution to create this core-halo spans $1.4$ Gyr with a temporal resolution of $\Delta t = 10^{-4}$ Gyr.

\begin{figure}
\includegraphics[width=8cm]{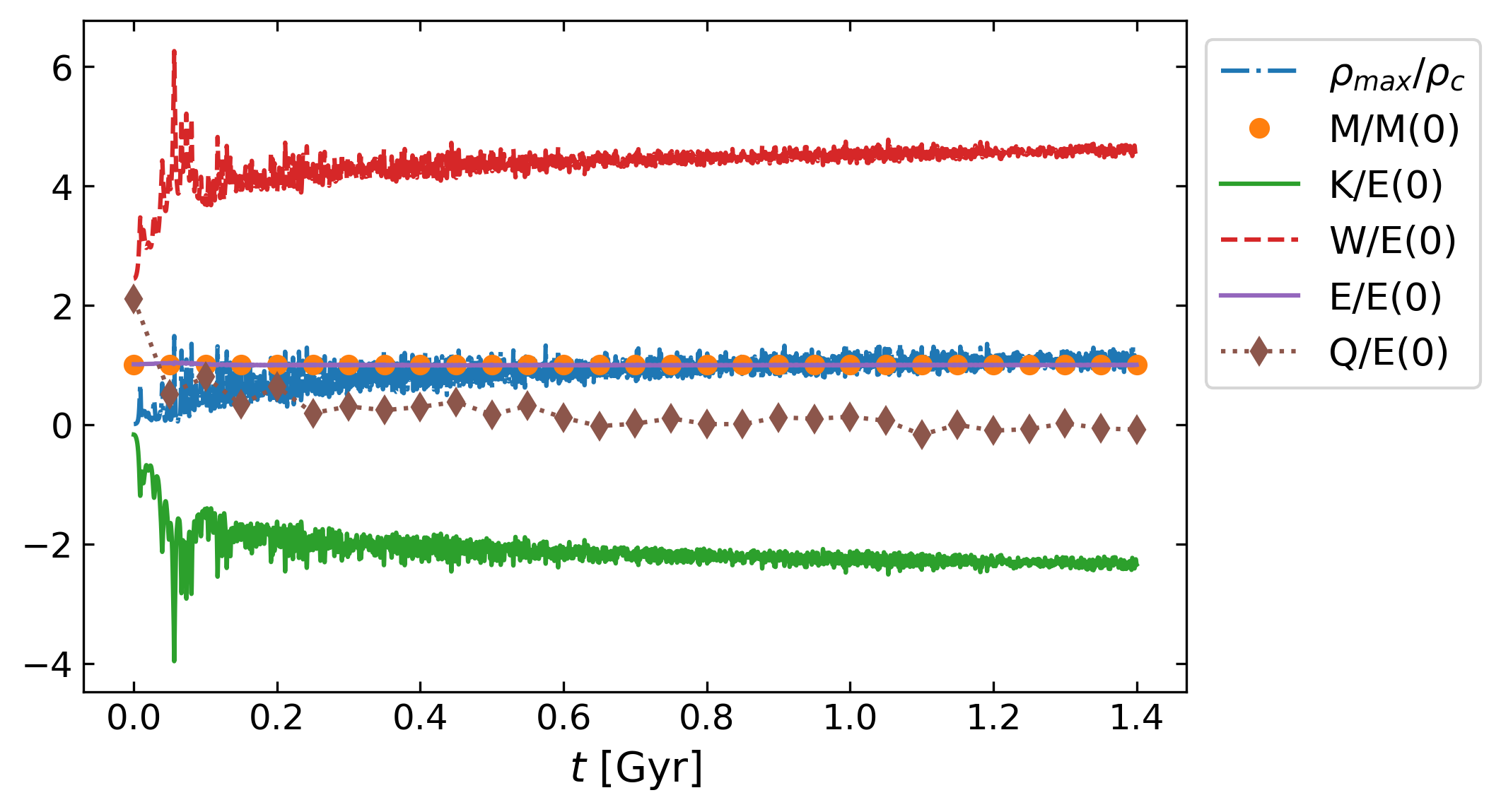}
\caption{Diagnostic showing the maximum density normalized with respect to the average core density, the kinetic energy ($K$), potential energy ($W$), total energy ($E = K + W$), and virial function ($Q = 2K + W$) normalized with respect to the initial total energy $E(0)$, as well as the total mass ($M$) normalized with respect to the initial mass. These parameters highlight the halo's relaxation process, with density oscillations around a mean value. The virial function indicates virialization ($Q \approx 0$), and the stability of total energy and mass demonstrates the unitarity of the numerical evolution.}
\label{fig:diagnosticID}
\end{figure}

In Figure \ref{fig:diagnosticID}, we show the diagnostics of the system, that includes the maximum density normalized with the average core density, the kinetic energy $K:=-\frac{\hbar^2}{2m}\int \Psi^*\nabla^2 \Psi \, d^3x$, the potential energy $W := \frac{1}{2}\int \rho V \, d^3x$, the total energy $E := K + W$, the virial function $Q := 2K + W$ normalized with respect to the initial value of the total energy $E(0)\approx -1.652\times10^{16} \text{M}_\odot \text{km}^2/\text{s}^2$, and the total mass $M = \int \rho \, d^3x$, in turn normalized with respect to the initial value $M(0) \approx 2.598\times10^11$ M${}_\odot$. These quantities demonstrate the relaxation of the halo, where the density oscillates around a specific value. The virial factor shows that the halo is virialized, i.e., $Q \approx 0$. The kinetic and potential energies oscillate around ta nearly constant value, and the total energy and mass indicate how the numerical method maintains them nearly constant. 

\begin{figure}
\includegraphics[width=8.5cm]{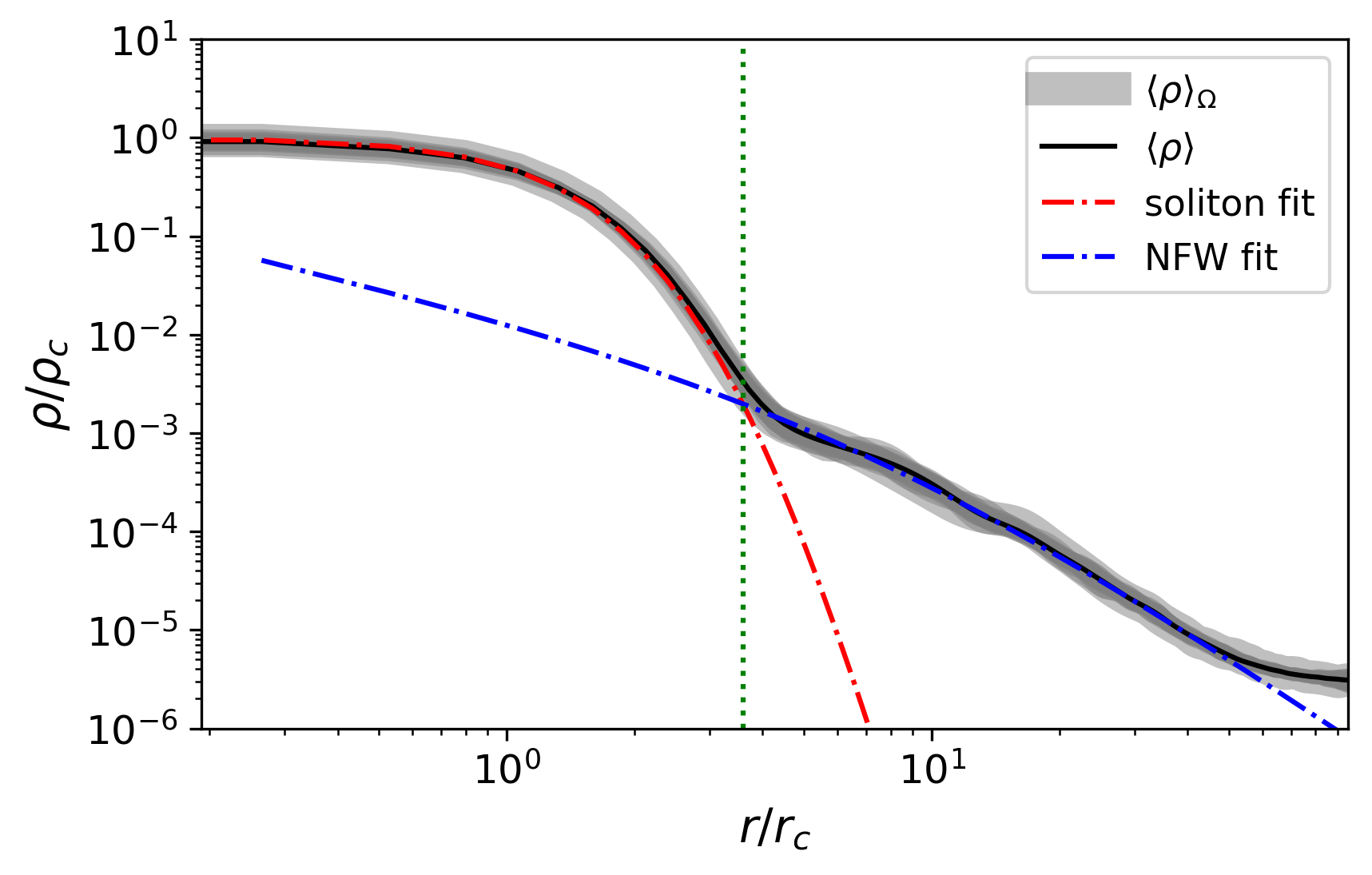}
\caption{Fitting of core and tail density profiles. The gray area represents many snapshots of the angle average density of the formed halo calculated using formula (\ref{eq:rhoaverage}) at  various times in the interval $[t_i,t_f]$, while the solid  black line represents the time average profile calculated with formula (\ref{eq:rhoaveragetot}), namely, the average of the gray lines. The red and blue lines correspond to the core and envelope fittings of the black curve, separately. The vertical line denotes the transition radius $r_t$, which separates the soliton core region ($r<r_t$) described by equation (\ref{eq:soliton}) from the tail region ($r>r_t$) described by the NFW profile (\ref{eq:NFW}).}
\label{fig:initialDensityavg}
\end{figure}

The density of the halo oscillates both in space and time around an average profile calculated as follows:

\begin{equation}
\langle \rho \rangle = \dfrac{1}{T} \int_{t_i}^{t_f} \langle \rho \rangle_{\Omega} \, dt,
\label{eq:rhoaverage}
\end{equation}

\noindent where

\begin{equation}
\langle \rho \rangle_{\Omega} = \dfrac{1}{4\pi} \int_{\Omega} \rho \, d\Omega,
\label{eq:rhoaveragetot}
\end{equation}

\noindent with $T = t_f - t_i$ being the time window over which a time-average is computed, with $t_i = 1.0$ and $t_f = 1.4$ Gyr, and $\Omega:=[0,\pi]\times[0,2\pi]$ being a solid angle over which the average is computed in space. This radial density profile generates a radial gravitational potential $\langle V \rangle$ through the Poisson equation (\ref{eq:Poisson}). This stationary density profile is characterized by a soliton core, enveloped by a profile that decays as $r^{-3}$, approximately described on average by an NFW profile, as shown in \cite{Mocz:2017wlg, periodicas}. In Figure \ref{fig:initialDensityavg}, the average density calculated according to this formula (\ref{eq:rhoaverage}) is presented, along with a fit using the empirical expression $\rho_{\text{CT}}(r) = \rho_{\text{core}}(r)\Theta(r_t - r) + \rho_{\text{tail}}(r)\Theta(r - r_t)$, where $\Theta$ is the step function, and $r_t$ is the transition radius between the soliton core and the envelope, where the core density is given by the formula \cite{Schive:2014dra}:

\begin{equation}
\rho_{\text{core}}(r) = \rho_c \left[ 1 + 0.091\left(\frac{r}{r_c}\right)^2\right]^{-8},
\label{eq:soliton}
\end{equation}

\noindent and the NFW profile in the tail region $r > r_t$ reads

\begin{equation}
\rho_{\text{tail}}(r) = \frac{\rho_s}{\frac{r}{r_s}\left(1 + \frac{r}{r_s}\right)^2},
\label{eq:NFW}
\end{equation}

\noindent with the fitting values $r_c = 0.2621$ kpc, $r_t = 0.9427$ kpc, and $r_s = 1.844$ kpc. These values satisfy the relation $r_t \approx 3.597 r_c$, which is approximately the value reported by \cite{Mocz:2017wlg}. The central density is obtained from the core radius through the expression

\begin{equation}
\rho_c = \dfrac{\hbar^2}{4\pi G m_B^2} \left(\dfrac{1.30569}{r_c}\right)^4 \approx 1.983 \times 10^7 \left(\dfrac{\text{kpc}}{m_{22}^2 r_c^4}\right) \text{M}_\odot,
\end{equation}

\noindent where $m_{22} = m_B \times 10^{22}$ eV, which can also be found with the numerical solution of the ground state. The density $\rho_s$ is obtained by assuming that the core-tail density is continuous at $r_t$. Thus, we have $\rho_s = \rho_{\text{core}}(r_t) \frac{r_t}{r_s} \left(1 + \frac{r_t}{r_s}\right)^2$, as in \cite{Tula2018}.

Also in Figure \ref{fig:initialDensityavg} we show a set of snapshots of the core-tail fitting density profiles, that superposed define a grey thick line in the plot. These snapshots reveal the core oscillation consistent with that presented in \cite{Veltmaat_2018}, which in turn is consistent with the fundamental quasinormal mode of the ground state solution of the SP system \cite{GuzmanUrena2004}, whose spectrum has been extended to higher order oscillation modes in \cite{Guzman2019}. On the other hand, the motion of the core with respect to the fixed reference frame defined by Figure \ref{fig:stationaryPeriod} is random, in consistency with \cite{Li_2021,Dutta_Chowdhury_2021}.

% ------------------------------
\subsection{Scenarios where test particles are to be studied}

So far we have constructed what will be our workhorse core-halo structure of FDM. Now, the motion of test particles will be studied in two clearly different scenarios: 

\begin{enumerate}
\item A first background test scenario for comparison, in which we assume that the space-time averaged density, the black line in Figure \ref{fig:initialDensityavg},  is time independent and spherically symmetric, sources the {\it stationary gravitational potential} $\langle V \rangle$. Test particles will be accelerated by the effects of the potential  $\langle V \rangle$.
\item A second scenario, in which the structure, even if nearly virialized is evolved in time, resulting in a highly kinetic  granular density in the halo and an oscillating core, which implies a spatial and time-dependent {\it gravitational potential} $V$. Test particles will be accelerated by the effects of the anisotropic and time-dependent potential  $ V $.
\end{enumerate}

In the first scenario one has to solve the equations of motion for test particles on a spherically symmetric stationary potential. This is a typical assumption when -for example- fitting galactic rotation curves, where granular structure is averaged out in order to simplify models as recently illustrated in \cite{IvanTulaChavanis2024}.

In the second scenario the structure evolves according to the SP system (\ref{eq:GP})-(\ref{eq:Poisson}), for which we simply continue the evolution of the core-halo for other additional 1.274 Gyr, and obtain $V$ at all times, that we use to integrate the equations of motion of test particles on the fly at the same time as the FDM evolves.

% ---------------------------------------------
% ----->     Analysis.    <-----
% ---------------------------------------------
\section{Equations of motion for test particles}
\label{sec:eqmotion}

We now write down the equations of motion of a test particle in each of the two scenarios.

% ------------------------------
\subsection{Scenario 1: Stationary Potential}

In this case, for the spatial and time average potential $\langle V\rangle$, which is spherically symmetric and time-independent, the equations of motion of a test particle take the form

\begin{equation}
\dfrac{d^2 \expval{\vec{x}_p}}{dt^2} = -\left.\nabla\expval{V}\right|_{\vec{x}=\expval{\vec{x}_p}} = -\left. \dfrac{d\expval{V}}{dr}\right|_{r=r_p} \dfrac{\expval{\vec{x}_p}}{r_p},
\label{eq:motion test particle avg}
\end{equation}

\noindent or equivalently

\begin{equation}
\dfrac{d^2 \expval{\vec{x}_p}}{dt^2} + \omega_{p}^2 \expval{\vec{x}_p} = 0,\label{eq:testparticlestationary}
\end{equation}

\noindent where 

\begin{equation}
\omega_{p}^2 := \frac{1}{r}\left. \frac{d\expval{V}}{dr}\right|_{r=r_p},
\label{eq:angularvelocitySP}
\end{equation}

\noindent and \(r_p = |\expval{\vec{x}_p}|\), in which \(\expval{\vec{x}_p}\) is used for the trajectory of the test particle in the stationary potential \(\expval{V}\), and also define \(\expval{\vec{v}_p} := \frac{d\expval{\vec{x}_p}}{dt}\) as its velocity. In this case, we have a well-known scenario: a particle under the influence of a central force. Here, the total specific energy $\expval{E_p} = \frac{1}{2}|\expval{\vec{v}_p}|^2 + \expval{V}|_{\vec{x}=\expval{\vec{x}_p}}$ and the total specific angular momentum $\expval{\vec{L}_p} = \expval{\vec{x}_p} \times \expval{\vec{v}_p}$ are conserved. The latter implies that the trajectory of the particle will occur only in the plane perpendicular to $\expval{\vec{L}_p}$ that we choose to be $z=0$. 

Now, the construction of circular trajectories in this potential will serve as a control case to compare with, because in the second scenario where the potential is time-dependent, trajectories may deviate from circular. Meanwhile, a particle in a circular orbit of radius $r_p$ has trajectory

\begin{equation}
\expval{\vec{x}_p} = r_p \left(\cos(2\pi \frac{t}{T_p}), \sin(2\pi \frac{t}{T_p}), 0\right),
\label{eq:solution test particle avg}
\end{equation}

\noindent where $T_{p} = 2\pi/\omega_{p}$ is the period in which a particle completes a circular orbit. 

% ------------------------------
\subsection{Scenario 2: Dynamic Potential}

In this scenario, a test particle experiences the force of gravity due to the time-dependent gravitational potential $V$. Consequently, the trajectory of the test particle evolves simultaneously with the dynamics of the FDM structure, following the equations of motion:

\begin{equation}
\dfrac{d^2\vec{x}_p}{dt^2} = -\left.\nabla V\right|_{\vec{x}=\vec{x}_p},
\label{eq:motion test particle general}
\end{equation}

\noindent where $\vec{x}_p$ represents the position of the test particle with velocity by $\vec{v}_p := \frac{d \vec{x}_p}{dt}$. It is important to note that, in this case, the specific total energy of the test particle $E_p = \frac{1}{2}|\vec{v}_p|^2 + \left.V\right | _{\vec{x}=\vec{x}_p}$, and the specific angular momentum $\vec{L}_p = \vec{x}_p\times\vec{v}_p$, are not conserved due to the time-dependency of the gravitational potential. This implies that the particle's trajectory will not occur only in the plane perpendicular to $\expval{\vec{L}_p}$ as in the stationary scenario.

% ------------------------------
\subsection{Initial conditions}

\begin{figure}
\includegraphics[width=8cm]{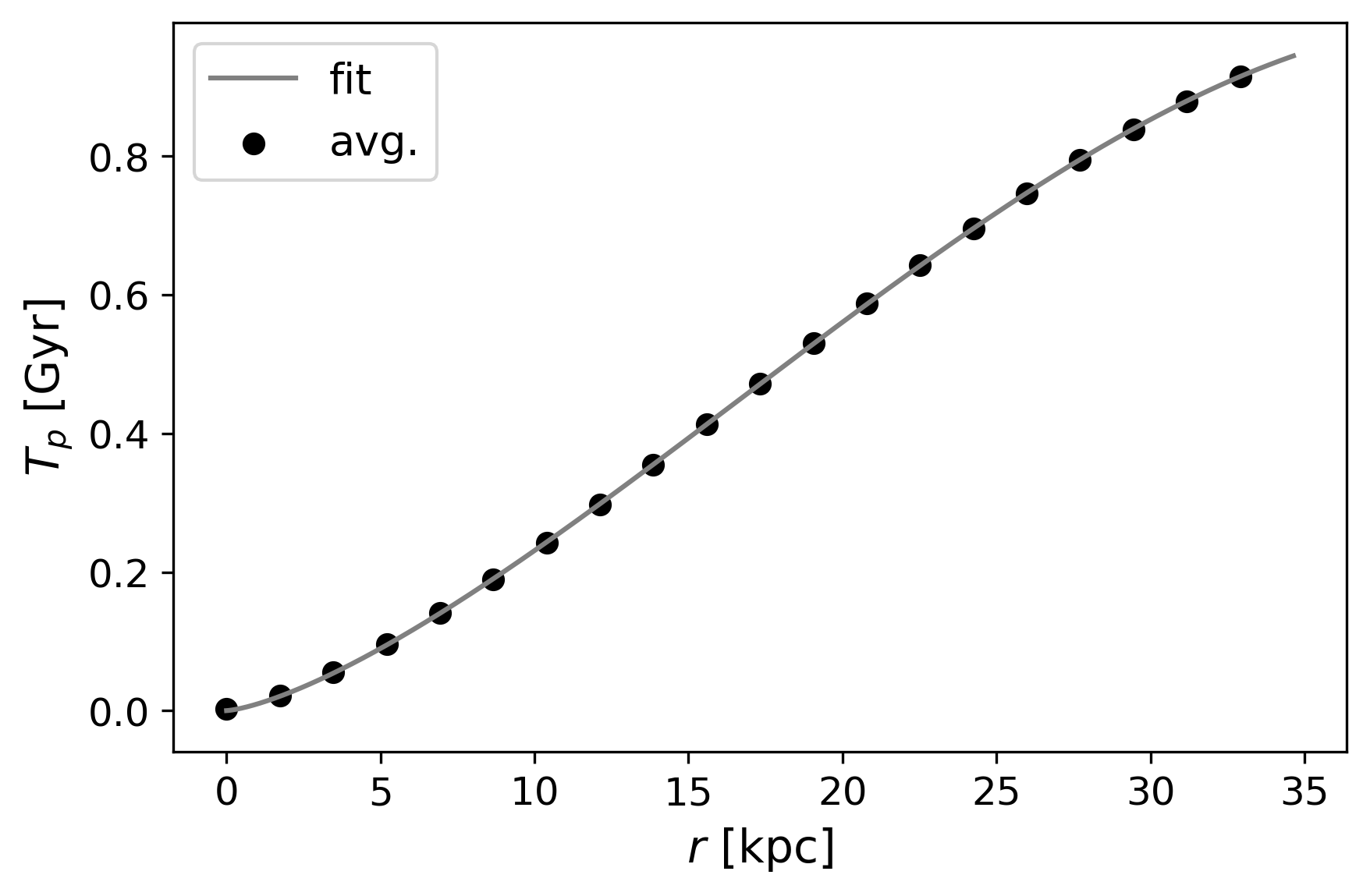}
\caption{Period in which a test particle completes a circular orbit around the system for the stationary potential $\expval{V}$ where the black points represent the numerical solution and the gray dotted line a fitting formula. This formula is used to  set initial conditions for circular trajectories for arbitrary $r$.}
\label{fig:stationaryPeriod}
\end{figure}

In order to study the difference of motion in the two scenarios, we will solve the equations of motion for a number of test particles whose initial conditions will correspond to circular trajectories at different radii. By solving numerically the equations (\ref{eq:testparticlestationary}) for different radius, and verifying formula (\ref{eq:solution test particle avg}), we find the orbital period as function of radius illustrated in Figure \ref{fig:stationaryPeriod} for the stationary potential $\expval{V}$. From this plot we extract the empirical formula

\begin{equation}
T_p = a_1 \left(\dfrac{r}{r_c}\right)^{1.392} \left(1 - a_2\left(\dfrac{r}{r_c}\right)^2\right),
\label{eq:stationaryPeriod}
\end{equation}

\noindent with the fitting parameters are $a_1 = 1.064\times10^{-3}$ and $a_2 = 1.668\times10^{-5}$.

Three sets of $N_p=100$ particles are defined. We use initial conditions for circular motion with initial  $\vec{x}_p(0) = r_p (1,0,0)$ and  velocity $\vec{v}_p = \frac{2\pi r_p}{T_p}(0,1,0)$, where the period $T_p$ is given by  formula (\ref{eq:stationaryPeriod}). We define three sets of  $N_p=100$ particles, the first set with uniformly distributed initial $x-$positions in the interval $r_p\in(0,r_0]$, where $r_0 = 92.99 r_c$. Observe that these particles have initial positions both inside the core and in the halo region. The second and third sets  have initial $x-$positions slightly deviated from those of the first set in such a way that for the second set of conditions $r_p$ is multiplied by a factor of $1.001$, and  for the third set by $1.01$. The idea is to later study the sensitivity to initial conditions of the trajectories in order to envisage possible chaotic behavior.

\subsection{Boundary conditions}

Usually, boundary conditions for ordinary differential equations of test particles are not important. However, when dealing with the potential required in Newton's second Law, which is confined to the numerical domain where the solution is known, the same boundary conditions as those in the SP system (\ref{eq:GP})-(\ref{eq:Poisson}) need to be applied, in our case periodic boundary conditions are employed. Consequently, a particle that exits the numerical domain through a given face renters through the opposite one, thereby maintaining consistency with the gravitational potential.

Finally, the integration of the motion equations above is carried out with a fourth-order Runge-Kutta method.

% ---------------------------------------------
% ----->     Results.    <-----
% ---------------------------------------------

\section{Motion of test particles}
\label{sec:soleqmotion}

\subsection{Scenario 1: Stationary Potential Case}

\begin{figure}
\includegraphics[width=8.5cm,height=3.75cm]{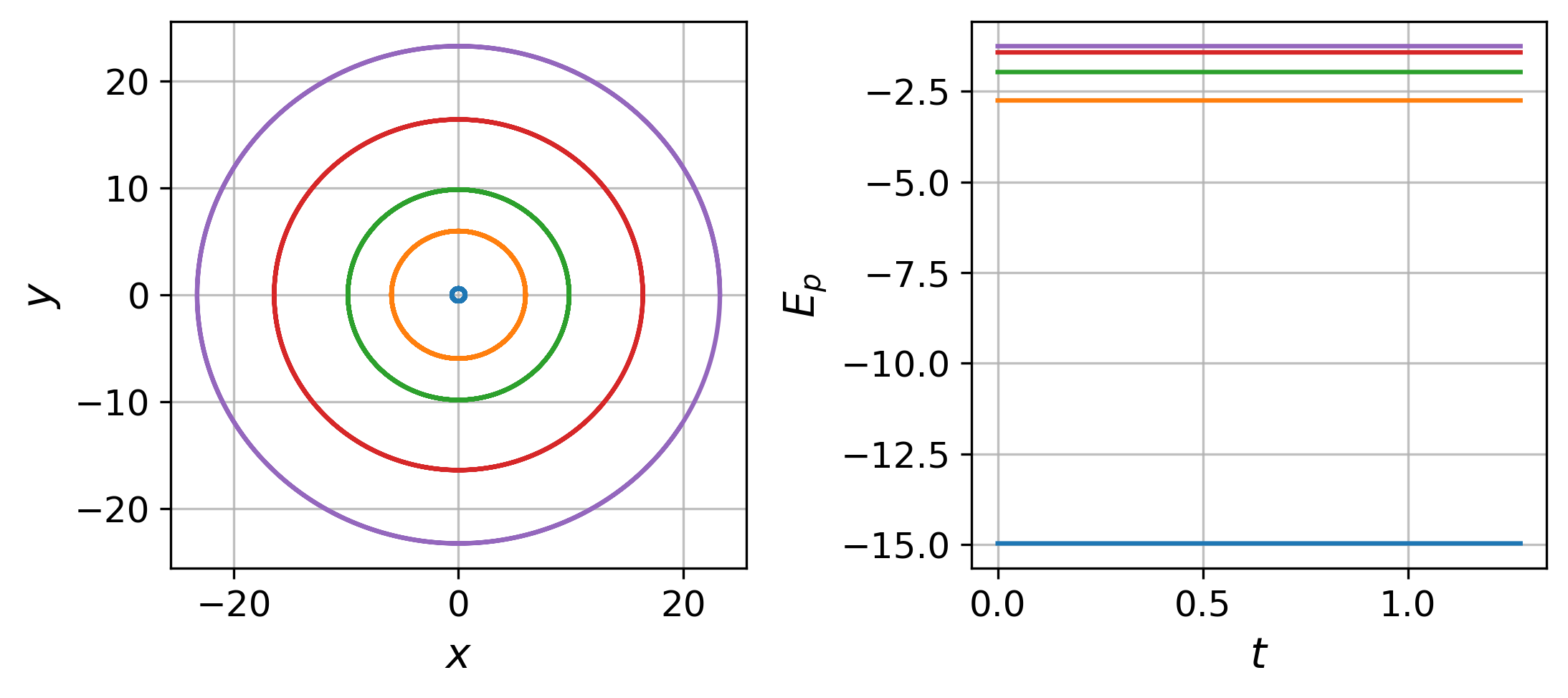}
\caption{On the left, trajectories of a sample of test particles in the stationary potential at the plane $z=0$, which are circular and thus consistent with the description of the initial conditions. On the right, the total energy as a function of time for the same sample of test particles, which illustrates the conservation of energy during the numerical integration of trajectories.}
\label{fig:stationaryCircularTrajec}
\end{figure}

In order to test that numerical integration of trajectories works fine, we solve motion equations (\ref{eq:testparticlestationary}) for the potential  $\expval{V}$ and compare with the exact solution (\ref{eq:solution test particle avg}). The results of numerical integration are consistent with the exact ones, according to which the particles; positions are given by 
$\expval{x_p} = r \cos\left[\omega_{r} t\right]$,  
$\expval{y_p} = r \sin\left[\omega_{r} t\right]$, 
$\expval{z_p} = 0$, where angular frequency and period for a particle starting a radius $r$ are respectively $\omega_r = \sqrt{\frac{1}{r}\frac{d\expval{V}}{dr}}$ and $T_r = 2\pi / \omega_r$. 
The results are exemplified in Figure \ref{fig:stationaryCircularTrajec} for a subsample of the whole set of initial conditions. On the left we show that  the trajectories are truly circular and on the right that the conservation of energy is satisfied within numerical precision, as the trajectories are maintained on the same path during 1.274 Gyr. This test illustrates for how long the numerical integration of trajectories maintains the energy, as well as the appropriate implementation of the interpolation of the test particles trajectories within the numerical grid where the variables of the SP system are defined. 

%----------------------
\subsection{Scenario 2: Dynamic Potential}

In this case we solve the motion equations (\ref{eq:motion test particle general}) on top of the evolution of the core-halo structure according to the system (\ref{eq:GP})-(\ref{eq:Poisson}) from which we extract $V$ at all times. We integrate trajectories for all initial conditions and track the particles' trajectories that were circular in the stationary potential case.

\begin{figure*}
\includegraphics[width=16cm]{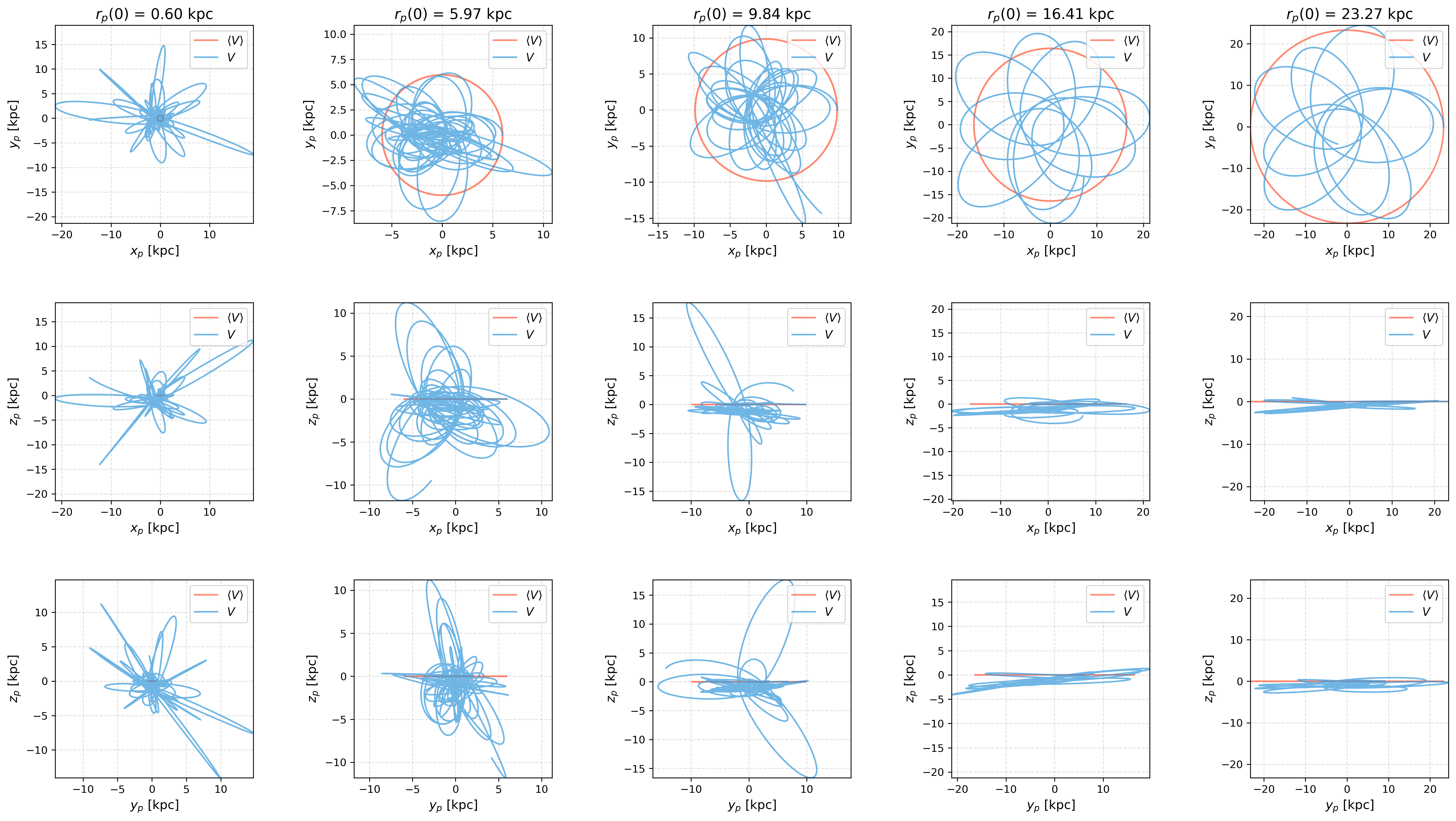}
\caption{Orange lines represent the trajectories of test particles under the influence of the stationary potential $\expval{V}$, for five initial conditions,  noticeable in the projection on the $xy-$plane. Blue lines represent the trajectories of test particles under the fully time-dependent potential $V$. The first, second, and third rows show the projections of the trajectories on the $xy$, $xz$, and $yz$ planes, respectively. Each column corresponds to the initial conditions $x_p(0) = 0.6000$, $6.560$, $12.53$, $18.49$, and $-15.27$ from left to right. The reference frame of these trajectories is the laboratory frame, where the box of the simulation is kept fixed.
}
\label{fig:Traject stationary vs dynamic}
\end{figure*}

Figure \ref{fig:Traject stationary vs dynamic} illustrates the projection of some test particles' trajectories on different planes, under the influence of the fully time-dependent potential $V$. They are compared with those for the stationary average potential $\expval{V}$. The trajectories exhibit an erratic behavior for small initial radius close and within the core, where they appear to be chaotic. For larger initial radii, away from the core, the trajectories seem to be more uniform, although they also deviate from circular paths.

\begin{figure}
\includegraphics[width=8cm]{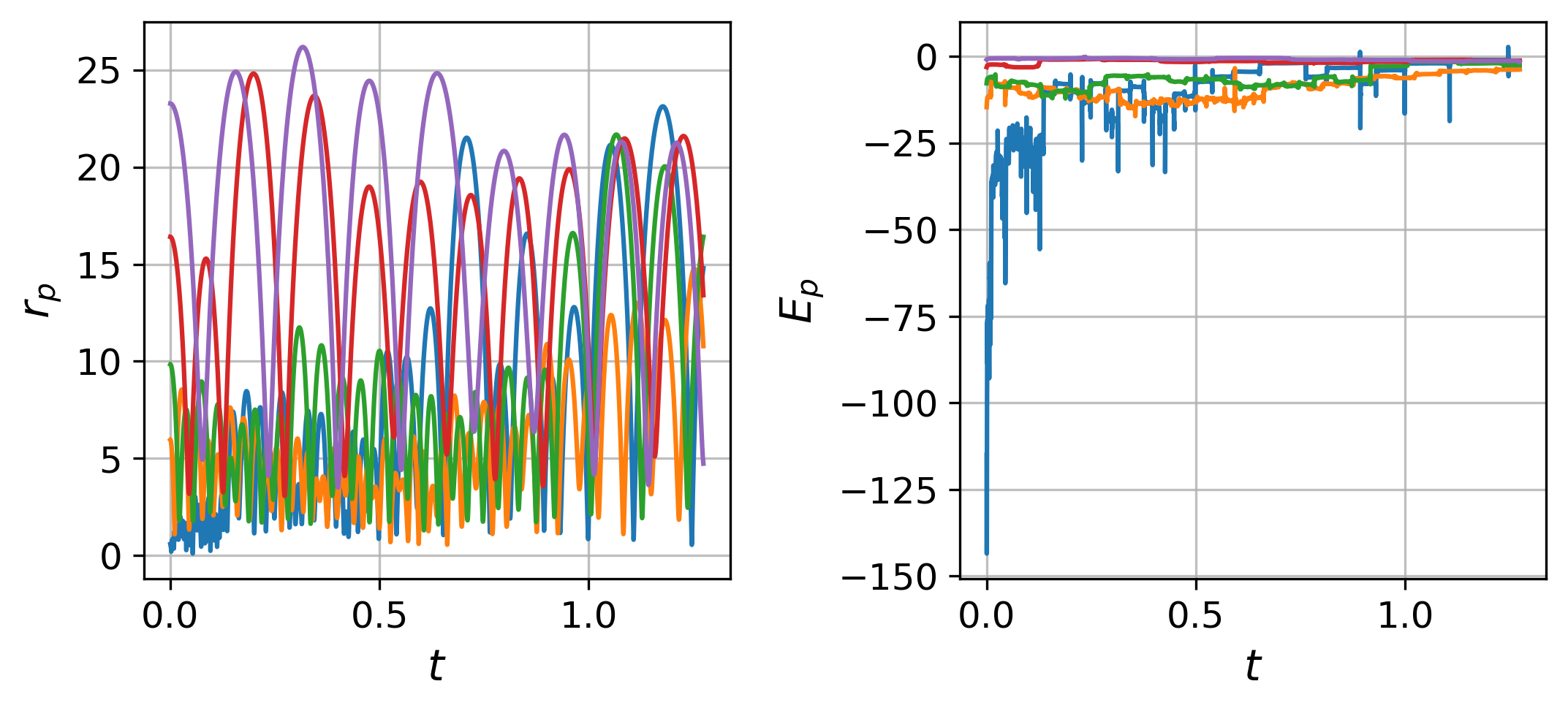}
\caption{On the left, the distance to the origin of test particles as function of time, in the case of the Dynamic potential $V$. These differ from the constant radius corresponding to the stationary case in Figure \ref{fig:stationaryCircularTrajec}. These differences arise because the dynamic potential accelerates the particles and  consequently, the energy is not conserved, as illustrated on the right.}
\label{fig:dynamic r and E}
\end{figure}

For a closer examination, Figure \ref{fig:dynamic r and E} on the left presents the radial distance of the particles $r_p = \sqrt{x_p^2 + y_p^2 + z_p^2}$ as function of time. This Figure illustrates how $V$ heats the particles with different modes and strength over time. At the right we draw the total energy of the particles, which is not conserved, and instead indicates how the potential injects energy to the trajectories and ``heats'' the motion of particles, an effect that happens even to Black Holes within FDM halos \cite{Boey2024}.

% -----------------------
\subsection{Chaotic trajectories}

The trajectory of a particle appears to be chaotic in the fully time-dependent potential case, exhibiting a markedly different behavior from the averaged stationary potential case. However, we will now demonstrate that the trajectory is chaotic in a more strict sense. A first-order dynamical system is said to exhibit chaos if (see e.g. \cite{ott2002chaos}):

\begin{enumerate}
\item It is a nonlinear system.
\item It is a system in at least three dimensions.
\item It is sensitive to initial conditions.
\end{enumerate}

\begin{figure*}
\includegraphics[width=16cm]{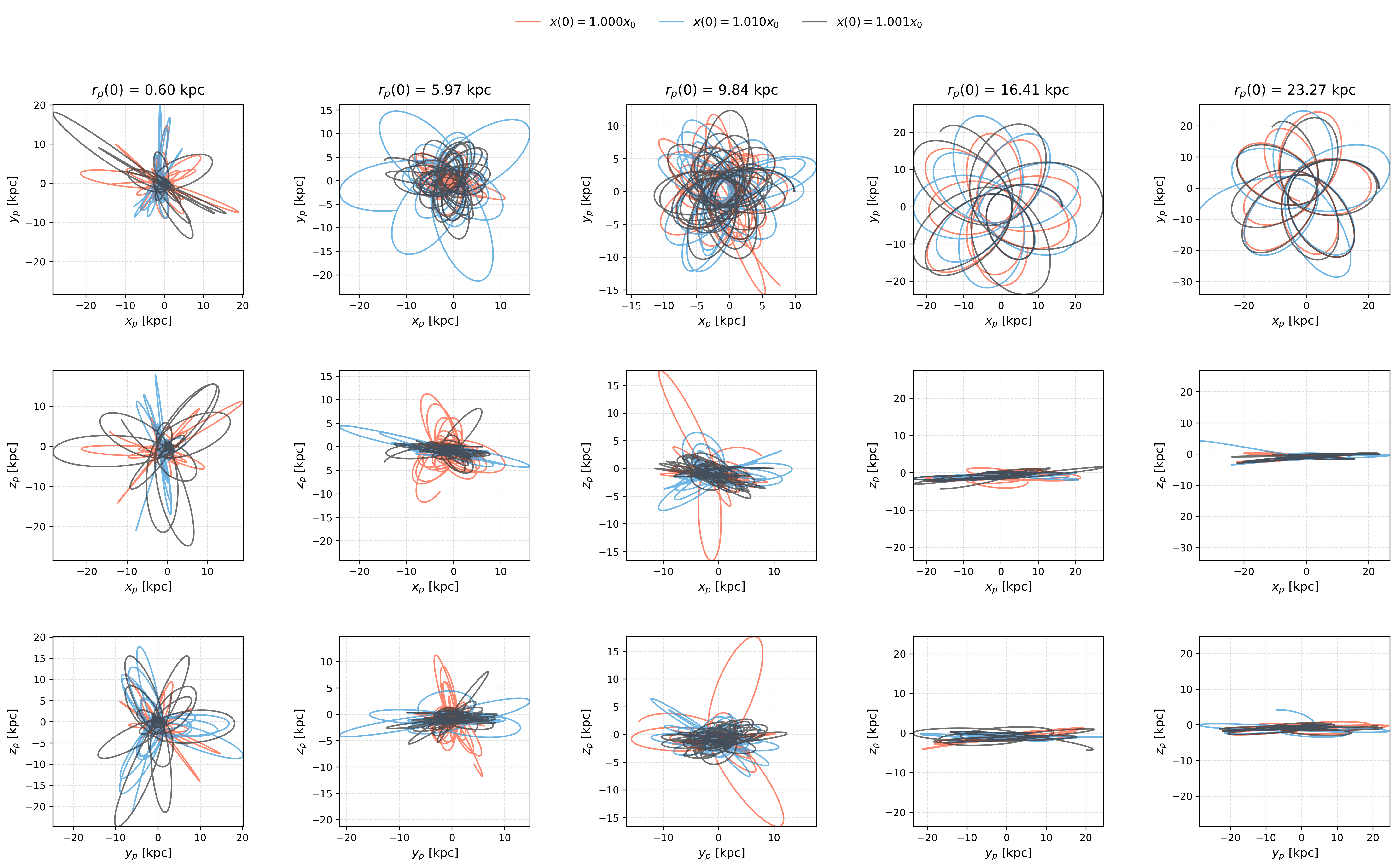}
\caption{Sensitivity to the initial position of the trajectory of five tests particle with three close initial conditions $x_p(0) = x_0$ in orange, $x_p(0) = 1.01 x_0$ in blue and $x_p(0) = 1.001 x_0$ in gray. The first, second, and third rows show the projections of the trajectories of the particles in the $xy$, $xz$, and $yz$ planes, respectively. Each column corresponds to the initial conditions $x_p(0) = 1.19$, $4.18$, $7.16$, $10.14$, and $13.42$ kpc from left to right. The reference frame of these trajectories is the laboratory frame, where the box of the simulation is kept fixed.}
\label{fig:ICsensitivity}
\end{figure*}

\noindent The first two conditions are satisfied for the test particle evolution equations, since the potential is nonlinear and the system of equations of motion of the test particle conforms a system of six first order equations in time. To verify the third condition, we use the trajectories of the three sets of initial conditions for particles defined above. The top, middle, and bottom rows of Figure \ref{fig:ICsensitivity} show the projections of five particle trajectories on the $xy$, $xz$, and $yz$ planes, respectively. Each column corresponds to a value of the initial conditions $x_p(0) = 1.19$, $4.18$, $7.16$, $10.14$, and $13.42$ kpc from left to right. The second and third sets have an initial $x$ initial coordinate, modified by a factor of 1.01 and 1.001 respectively. Notice that for the first three columns, corresponding to initial positions closer to the core, the trajectories are very sensitive to the initial condition. For the last two columns corresponding to positions farther from the core, the trajectories at least remain close to the plane where the circular trajectory of the initial conditions, although the trajectories continue to be sensitive to initial conditions and are not circles. This illustrates how the motion and oscillations of the core, where the potential gradient is more dynamical and steeper, impact on the trajectories more importantly than in the halo region.

\begin{figure}
\includegraphics[width=8cm]{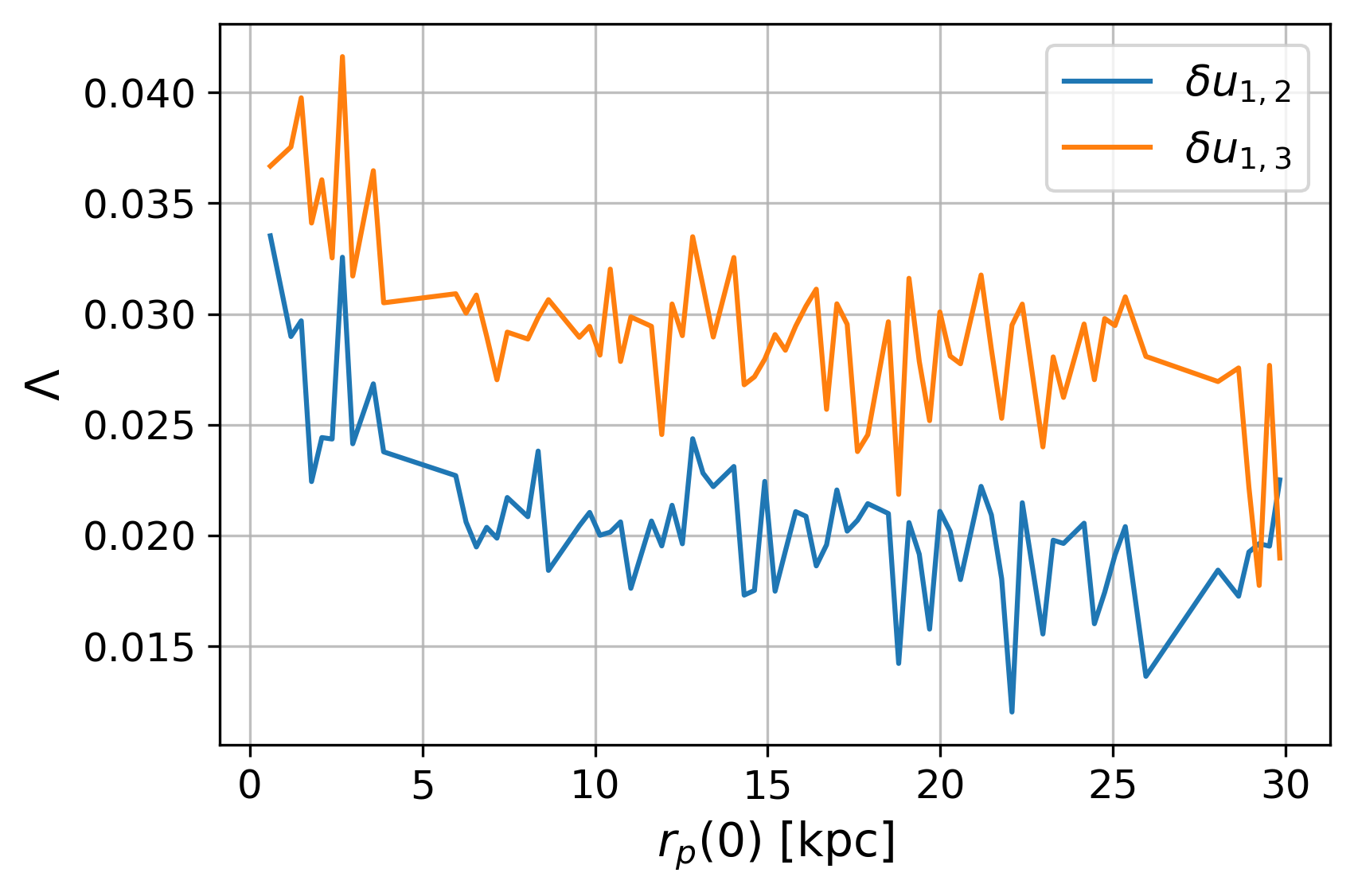}
\caption{Maximal Lyapunov exponent as a function of the initial $x-$coordinate of test particles. Yellow line corresponds to deviations between initial conditions $x_p(0) = x_0$ and $x_p(0) = 1.01 x_0$, whereas blue line corresponds to deviations between initial conditions $x_p(0) = x_0$ and $x_p(0) = 1.001 x_0$.}
\label{fig:lyapunov}
\end{figure}

In order to check whether there is an exponential deviation between trajectories initially very close, we calculate the Maximum Lyapunov Exponent (MLE) defined as

\begin{equation}
\Lambda = \lim_{t\to\infty} \lim_{|\delta \vec{u}(0)|\to 0} \dfrac{1}{t}\log\left(\dfrac{|\delta \vec{u}(t)|}{|\delta \vec{u}(0)|}\right),
\label{eq: Lyapunov exponent}
\end{equation}

\noindent where $\vec{u} = (\vec{x}_p, \vec{v}_p)$ is the phase space position of the particle, which is the state of the six-dimensional dynamical system, and $\delta\vec{u}$ is the displacement of the two trajectories that are being compared, with initial separation $\delta\vec{u}(0)$. However we do not know the solution in the continuous domain, for this reason we approximate the MLE exponent as

\begin{equation}
\Lambda \approx \dfrac{2}{t_f}\log\left(\dfrac{|\delta \vec{u}(t_f/2)|}{|\delta \vec{u}(0)|}\right),
\end{equation}

\noindent  where $t_f=1.274$Gyr. Figure \ref{fig:lyapunov} shows the MLE for the displacement vectors between the first and the second set of initial conditions, and between the first with the third set. Notice two things, first that irrespective of the initial displacement in the initial conditions the MLE is always positive, indicating that chaos is present across the entire numerical domain. Second, the separation of close trajectories is most pronounced in a region near and within the core.

% ---------------------------------------------
% ----->     Conclusions.    <-----
% ---------------------------------------------
\section{Conclusions}
\label{sec:comments}

We have shown the effects of the anisotropy and time-dependence of a core-halo FDM structure on test particles 
integrated in two different scenarios. The first one assumes that the gravitational potential $\langle V \rangle$ is spherically symmetric and stationary, obtained from the angule and time averaged gravitational potential $V$ that is non-symmetric and non-stationary. The second scenario assumes test particles are subject to $V$ during the integration in a time-window of $\sim$ 1.6Gyr. 

The initial conditions of test particles correspond to circular trajectories in the laboratory reference frame defined in Figure \ref{fig:initialDensity}, which we verified remain circular when using $\langle V \rangle$. The same trajectories are integrated using $V$ and do not remain circular, not even on the same plane. Instead they show erratic behavior, which is more evident for trajectories within or near the core than in the halo region.

In order to quantify the sensitivity to initial conditions of trajectories, we integrated test particle trajectories with close initial positions and determined the Lyapunov exponent of their departure. The result is that it is always positive, which indicates that trajectories are chaotic in the whole domain, with bigger exponents near the core. This is consistent with the random walk motion of the core described in \cite{PhysRevLett.124.201301,Dutta_Chowdhury_2021}, and to the oscillations of the core during the evolution of a core-halo \cite{Li_2021,Veltmaat_2018,IvanTulaChavanis2024} in non symmetric simulations, that are in agreement with the low order quasinormal  mode oscillation found in \cite{GuzmanUrena2004} and later extended in \cite{Guzman2019}. The core's random motion and oscillation implies a more dynamical gravitational potential where gradients are steeper, and test particles are heated more \cite{Dutta_Chowdhury_2023}, which leads to a more noticeable departure of trajectories from circular.

A direct implication of this result is, that trajectories of test particles obtained when considering the FDM structure to be the one given by the stationary and spherically symmetric core-halo formulas, which are actually averages of the anisotropic and time-dependent configuration, are very different from those obtained considering the true one, namely the non-averaged, anisotropic, time-dependent configuration. This should raise a warning on the use of oversimplified spherically symmetric smooth FDM distribution in phenomenological studies.

The collective behavior of particles, for example that of a gas on top of an FDM core-halo structure, may reveal new correlations and possibly a more stationary collective behavior, like those obtained for collective systems in (e.g. \cite{El_Zant_2019,Bar_Or_2019}). The study that includes evolution of both FDM and gas is the subject of current analysis \cite{FDMplusGAS}.

\section*{Acknowledgments}
We thank Jens Niemeyer and two anonymous Referees for useful comments. 
Iv\'an Alvarez receives support within the CONACyT graduate scholarship program under the CVU 967478. 
This research is supported by grants CIC-UMSNH-4.9, Laboratorio Nacional de C\'omputo de Alto Desempe\~no Grant No. 1-2024 and 5-2025, CONAHCyT Ciencia de Frontera 2019 Grant No. Sinergias/304001.

% -------------------------------------------------------
% -----     REFERENCES     ----------
% -------------------------------------------------------

\bibliography{BECDM}

\end{document}